\documentclass[pre,twocolumn,showpacs]{revtex4}

\usepackage{graphicx}%
\usepackage{dcolumn}
\usepackage{amsmath}

\makeatletter
\def\btt#1{\texttt{\@backslashchar#1}}%
\DeclareRobustCommand\bblash{\btt{\@backslashchar}}%
\makeatother

\begin{document}


\title{$N$-dimensional nonlinear Fokker-Planck equation with
time-dependent coefficients}
\author{L. C. Malacarne, R. S. Mendes and I. T. Pedron}
\affiliation{Departamento de F\'\i sica, Universidade Estadual de Maring\'a, \\
Av. Colombo 5790, 87020-900 Maring\'a-PR, Brazil}

\author{E. K. Lenzi}
\affiliation {Centro Brasileiro de Pesquisas F\'\i sicas, R. Dr.
Xavier Sigaud 150, 22290-180 Rio de Janeiro-RJ, Brazil }

\date{\today}

\begin{abstract}
An $N$-dimensional nonlinear  Fokker-Planck equation is
investigated  here by considering the time dependence of the
coefficients, where drift-controlled and source terms are present.
We exhibit the exact solution based on the generalized gaussian
function related to the Tsallis statistics. Furthermore, we show
that a rich class of diffusive processes, including normal and
anomalous ones, can be obtained by changing the time dependence of
the coefficients.
\end{abstract}

 \maketitle

Anomalous diffusion processes appear in a large class of systems
in several contexts. Some illustrative examples are diffusion in
plasmas\cite{Berryman},  relative diffusion in turbulent
media\cite{ProcacciaPRA}, CTAB micelles dissolved in salted
water\cite{Ott}, surface growth and transport of fluid in porous
media\cite{spohn93}, two dimensional rotating flow\cite{Solomon},
subrecoil laser cooling\cite{Bardou}, diffusion on
fractals\cite{Stephenson}, anomalous diffusion at liquid
surfaces\cite{Bychuk}, diffusion in linear shear flows\cite{Jou1},
enhanced diffusion in active intracellular transport\cite{Caspi},
particle diffusion in a quasi-two-dimensional bacterial
bath\cite{Wu}, and spatiotemporal scaling of solar surface
flows\cite{Lawrence}, among others.

The existence of the anomalous diffusion and its ubiquity has
motivated, in particular,  the analytical study by considering
nonlinear\cite{plastino95,tsallis96,lisa,lisa2,Jou2,plastino24,Malacarne}
and fractional\cite{Metzler,Bologna} Fokker-Planck equations,
spatial dependence of the diffusion
coefficient\cite{Malacarne,lisa2,richardson26,shaughnessy84}, and
temporal dependence of the drift term\cite{plastino24,lillo}.

In this brief report, we consider an $N$-dimensional nonlinear
Fokker-Planck equation  incorporating  the time dependence in
every coefficient, where drift-controlled (external force) and
source terms are present. In our exact solution, a rich class of
anomalous behaviors can arise by choosing appropriate time
dependence of the coefficients. These results indicate that the
possible anomalies in diffusive processes can appear as a
consequence of different causes. This paper contains, essentially,
the main results obtained in Refs.
\cite{plastino95,tsallis96,plastino24} and \cite{lillo} as special
cases.

In order to investigate, in an exact way, a large class of
anomalous diffusions, let us consider the $N$-dimensional
nonlinear Fokker-Planck equation
\begin{widetext}
\begin{equation}
\frac{\partial }{\partial t}\hat{\rho} ({\bf {r}},t)={\cal D}
(t)\nabla ^{2}\left[ \hat{\rho} ( {\bf {r}},t)\right] ^{\nu }-{\bf
\nabla}\cdot \left[ { \bf F}({\bf {r}} ,t)\hat{\rho} ({\bf
{r}},t)\right] - \alpha (t)\hat{\rho}({\bf {r}},t), \label{S1}
\end{equation}
\end{widetext}
where we incorporate the time dependence in the external force,
\begin{equation}
{\bf F}({\bf {r}},t)= {\bf k}_1 (t) + k_2(t) {\bf r},  \label{f1}
\end{equation}
in the diffusion coefficient ${\cal D} (t)$, and  in the source
term $\alpha (t)$. In particular, we consider the case where the
space constant term in ${\bf F}({\bf {r}},t)$ can be taken with
different coefficients. This situation is useful, for example,  to
study diffusion in the gravitational field [${\bf
k}_1=(0,0,k_{1z})$]. The harmonic potential is considered
isotropic.

The source term in Eq.(\ref{S1}) can be removed by an appropriate
change in the solution:
\begin{equation}\label{ra}
\hat{\rho} ({\bf r},t)= \exp\left[ -\int_0^t
\alpha(t'){d}t'\right] {\rho}({\bf r},t).
\end{equation}
This way, ${\rho}({\bf r},t)$ obeys the equation
\begin{equation}
\frac{\partial }{\partial t}{\rho}({\bf r},t)=D(t)\nabla
^{2}\left[ {\rho}({\bf r},t)\right] ^{\nu }-{\bf \nabla}\cdot
\left[ {\bf F}({\bf {r}} ,t){\rho}({\bf r},t) \right] ,
\label{S1b}
\end{equation}
with $D(t)={\cal D}(t)
\exp\left[{(1-\nu)\int_0^t\alpha(t'){d}t'}\right]$. Thus, Eq.
(\ref{S1b}) has the same structure of Eq. (\ref{S1})   without the
source term, but with an additional time dependence of the
diffusion coefficient. Observe that this additional time
dependence of $D(t)$ is induced by the nonlinear term $\rho^\nu$,
disappearing when $\nu=1$.

In order to obtain an exact solution for Eq. (\ref{S1b})  with the
external force term given by Eq. (\ref{f1}),  we are going to
employ the {\it ansatz}
\begin{equation}
\rho ({\bf {r}},t)=\frac{1}{Z(t)}\left[1-(1-q)\beta (t)\left( {\bf {r}}-{\bf {r}%
}_{0}(t)\right) ^{2}\right]^{1/(1-q)} \label{S2}
\end{equation}
if  $ 1-(1-q)\beta (t) ({\bf r}- {\bf r}_0)^2  \geq 0$, and $\rho
({\bf r},t)=0$ if $ 1-(1-q)\beta (t) ({\bf r}- {\bf r}_0)^2 <0$
(cut-off condition). We would like to remark that Eq. (\ref{S2})
can be justified via dimensional analysis and related to the
renormalization group theory\cite{GoldenfeldBook}. Furthermore,
this {\it ansatz}  can also be  connected with Tsallis
statistics\cite{1a,Goldenfeld}. In addition, Eq. (\ref{S2})
reduces to the Gaussian when $q \rightarrow 1$. In fact, by
defining the function $\exp_q (-x^2) \equiv [1-(1-q)
x^2]^{1/(1-q)}$ if $1-(1-q) x^2\geq 0$, and $\exp_q (-x^2) \equiv
0$ if $1-(1-q) x^2 < 0$  as a $q$-gaussian, we obtain the usual
gaussian function by taking the limit $q\rightarrow 1$. Eq.
(\ref{S2}) is a solution of Eq. (\ref{S1b}) when $\nu =2-q$ and
the time dependence of $\beta (t)$, $Z(t)$, and ${\bf {r}}_{0}(t)=
\sum_{i=1}^N x_{0i} (t) {\bf e}_i$ are ruled by the following
system of equations:
\begin{eqnarray}
\frac{1}{Z}\frac{d Z}{d t} &=&2(2-q) N D \beta
Z^{-1+q}- N k_{2} ,\label{A1} \\
\frac{1}{\beta }\frac{d \beta }{d t} &=&-4(2-q) D Z^{-1+q}\beta
+2k_{2} , \label{A2}
\end{eqnarray}
and
\begin{equation}
 \frac{d {x}_{0i}}{d t} = k_{1i}-k_{2} x_{0i}.
\label{A3}
\end{equation}
Note that  Eqs. (\ref{A1}) and (\ref{A2}) are nonlinear, Eq.
(\ref{A3}) is independent of $\beta(t)$ and $Z(t)$, and
$k_{1i}(t)$ does not appear in the nonlinear coupled differential
equations for $\beta(t)$ and $Z(t)$. Furthermore, the spatial
independent term in the external force only affects the time
dependence of $x_{0i}$. So, Eq. (\ref{A3}) leads to
\begin{equation}\label{x0}
  x_{0i} (t) = e^{-\mu(t)} \left[ x_{0i} (0) + \int_0^t k_{1i} (s)
e^{-\mu (s)} ds\right],
\end{equation}
where $\mu (t) = \int_0^t k_2 (s) ds$. For example, in a
3-dimensional space with the presence of an isotropic
time-independent harmonic potential and the gravitational field,
{\it i.e.},  $k_2(t) = k_2= \mbox{constant}$, $k_{1x}=k_{1y}=0$
and $k_{1z} (t)= k_{1} =\mbox{constant}$, we get $x_0 (t) = x_0
(0) e^{-k_2 t}$, $y_0 (t) = y_0 (0) e^{-k_2 t}$, and
\begin{equation}\label{gravitational}
z_0 (t) =\left[ z_0 (0)+ \frac{k_1}{k_2} \left(e^{k_2 t} -1\right)
\right] e^{-k_2 t}.
\end{equation}

The solution for the nonlinear coupled equations for $Z(t)$ and
$\beta (t)$, Eqs. (\ref{A1}) and (\ref{A2}), is given by

\begin{equation}\label{SolA1}
  Z (t) = Z_0 \left[ 1-\frac{c_1}{N}
f(t)\right]^{N/c_1}
\end{equation}
and
\begin{equation}\label{SolA2}
  \beta (t) = \beta_0 \left[ 1-\frac{c_1}{N}
f(t)\right]^{-2/c_1},
\end{equation}
with $c_1= 2+N(1-q)$, $\beta_0 =\beta(t=0)$, $Z_0= Z(t=0)$ and
\begin{eqnarray}\label{ft}
  & & f(t) = e^{-c_1 \mu(t)} \times \nonumber \\ & &\int_0^t \left[ \frac{}{} N k_2
(s) - 2N(2-q) \beta_0 Z_0^{q-1} D(s)\right] e^{c_1
\mu(s)}ds.\nonumber \\
\end{eqnarray}

It is usual to identify normal diffusion  process by a linear
growth in time of the variance $\sigma^2\equiv <({\bf r}- {\bf
r_0})^2>$. Other time dependences on $\sigma^2$ are commonly
related as anomalous diffusion, for instance,  superdiffusive,
subdiffusive, exponentially diffusive, and localized. In our
study, we obtain a large class of diffusive processes which
include these examples. In addition, $\sigma^2$ can have different
behaviors for small and large times, enabling the description of
a rich structure of diffusion regime. Here we are going to
analyze, as an illustration, a set of  representative kinds of
these anomalous behaviors in the asymptotic regime taking some
specific dependence of the coefficients into account.  In this
direction, from Eqs. (\ref{ra}) and (\ref{S2}), we investigate the
asymptotic temporal behavior of the variance
\begin{equation}\label{variance}
  \sigma^2 = \frac{\int ({\bf r}- {\bf r}_0)^2 \hat{\rho} ({\bf
r},t) d^N{\bf r}}{ \int \hat{\rho} ({\bf r},t) d^N {\bf r}} =C
(q,N) \beta ^{-1},
\end{equation}
where $C(q,N)$ is a constant depending only on $q$ and $N$. Note
that the convergence of the integral  for $q>1$ in  $\sigma^2$
imposes a restriction over the parameters: $2+N(1-q)>2(q-1)$. This
implies that $c_1$ is a positive constant for all $q$ values.

Firstly, let us consider the case without the source term, with
the diffusion coefficient constant, and with the time dependence
of the harmonic external force given by $k_2(t)= k ~ t^{-b}$. From
Eq. (\ref{ft}), we obtain

\begin{equation}\label{case1}
  f(t) \sim \left[1 - e^{-c_1 k t^{1-b}/(1-b)}\right] +  \left[c_2 e^{-c_1
k t^{1-b}/(1-b)} - c_3 t^b \right]
\end{equation}
for large $t$ and $b<1$, where $c_2$ and $c_3$ are constants which
depend on $N,k,q$, and $b$. Again, for large t and $b=1$, we have
$f(t)\sim t^{-2k} - c_2 t^{-2k} (t^{1+2k}-1)/(1+2k)$, and finally,
for $b>1$, we get $f(t)\sim t$. By using the fact that the mean
square displacement is given by $\sigma^2 \sim f(t)^{2/c_1}$,
several asymptotic behaviors can be obtained. We summarize in Tab.
(\ref{table1}) the possible behaviors related to the above
asymptotic results. When we restrict our analysis to the
one-dimensional linear case ($\nu=1$), this drift-controlled
anomalous diffusion contains  the results given in \cite{lillo} as
a particular case.


\begin{table}
\caption{ Large time behavior of $\sigma^2 \sim f(t)^{2/c_1}$ for
$\alpha(t)=0$, $D(t)={\cal D}(t)=D_0$, and $k_2(t) =k t^{-b}$,
where $c_1=2+N(1-q)>0$.  \label{table1}}
\begin{ruledtabular}
\begin{tabular}{cccc}
  $b$ & $k$ & $\sigma^2 (t) $  & description \\ \hline \hline
  $b$ & $k=0$  & $t^{2/c1}$ & $c_1$-diffusive\footnotemark[1] \\ \hline
  $b=0$ & $k>0$ & $(1-e^{-c_1 kt})^{2/c_1}$ & Ornstein-Uhlenbeck
\\ \hline
  $b=0$  & $k<0$ & $e^{2 |k|t}$ & exponentially diffusive \\ \hline
  $0<b<1$ & $k>0$ & $t^{2b/c_1}$ & $c_1$-diffusive\footnotemark[2]
\\ \hline
  $0<b<1$ & $k<0$ & $e^{2 |k|t^{1-b}}$ & less than \\ & & & exponentially diffusive \\ \hline
  $b<0$ & $k>0$  & $1/t^{2|b|/c_1}$& localized \\ \hline
  $b<0$ & $k<0$  & $e^{2|k|t^{1-b}}$ & more than \\ & & & exponentially diffusive \\ \hline
  $b=1$ & $k>-1/2$ & $t^{2/c1}$ & $c_1$-diffusive\footnotemark[1] \\ \hline
  $b=1$ & $k=-1/2$ & $(t \ln t)^{2/c_1}$ & log divergent \\
\hline
  $b=1$ & $k<-1/2$ & $t^{2|k|}$ & superdiffusive \\\hline
  $b>1$ & $k$ & $t^{2/c1}$ & $c_1$-diffusive\footnotemark[1] \\
\end{tabular}
\end{ruledtabular}
\footnotetext[1]{The process is superdiffusive for $c_1<2$, normal
for $c_1=2$, and subdiffusive for $c_1>2$.} \footnotetext[2]{The
process is superdiffusive for $c_1< 2b$, normal for $c_1=2b$, and
subdiffusive for $c_1>2 b$.}
\end{table}


Consider now the nonlinear diffusion equation with neither  source
nor linear external force, but with the time dependence of the
diffusion coefficient given by $D(t)= {\cal D}(t)=D_0 t^d$. In
this case, for  large $t$, Eq. (\ref{ft})  leads to
\begin{equation}\label{case2}
  f(t) \sim t^{d+1}.
\end{equation}
Since  $\sigma^2 \sim f(t)^{2/[2+N(1-q)]}$, there is a competition
between the parameters $q$ and $d$ to define the diffusion
regimes. Tab. (\ref{table2}) contains a summary of these regimes.


\begin{table}
\caption{ Large time behavior of $\sigma^2 \sim f(t)^{2/c_1}$ for
$\alpha(t)=0$, $D(t)={\cal D}(t)=D_0 t^d$, and $k_2(t) =0$, where
$c_1=2+N(1-q)>0$.  \label{table2}}
\begin{ruledtabular}
\begin{tabular}{cccc}
  $d$ & $q$ & $\sigma^2 (t) $  & description \\ \hline \hline
  $d=0$ & $q=1$  & $t$ & normal \\ \hline
  $d=0$ & $q<1$ & $t^{2/[2+N(1-q)]}$ & subdiffusive \\\hline
  $d=0$ & $q>1$ & $t^{2/[2+N(1-q)]}$ & superdiffusive \\ \hline
  $d<0$ & $q=1$ & $t^{1-|d|}$ & subdiffusive \\ \hline
  $d>0$ & $q=1$ & $t^{1+d}$& superdiffusive \\ \hline
  $d>N(1-q)/2$ & $q$  & $t^{2(1+d)/[2+N(1-q)]}$& superdiffusive
\\ \hline
  $d<N(1-q)/2$ & $q$  & $t^{2(1+d)/[2+N(1-q)]}$ & subdiffusive \\
\end{tabular}
\end{ruledtabular}
\end{table}

Another possible situation is to take the nonlinear diffusion
equation with a time-dependent source term, $\alpha (t) =\alpha_0
t^a$, with time-independent diffusion constant, ${\cal D}(t)=D_0$,
and without linear external force, $k_2 (t)=0$. For $\nu=1 ~
(q=1)$ the source term does not affect the diffusion regime. On
the other hand, for $\nu \neq 1 ~(q\neq 1)$  and large $t$, Eq.
(\ref{ft}) gives
\begin{equation}\label{case3}
  f(t) \sim t^a \exp\left[ \frac{(q-1)\alpha_0 t^{1+a}}{1+a}\right]
\end{equation}
when $a>-1$. For $a=-1$, we have $f(t) \sim t^{(q-1)\alpha_0 +1}$,
and for $a<-1$, Eq. (\ref{ft}) reduces to $f(t)\sim t$. Tab.
(\ref{table3}) gives us a summary of the above behaviors. To
conclude our observations about the anomalous diffusion induced by
the time-dependent coefficients, we stress that the investigation
of a more complex time dependence of the coefficients can be
reduced to the analysis of Eq. (\ref{ft}).


\begin{table}
\caption{ Large time behavior of $\sigma^2 \sim f(t)^{2/c_1}$ for
$\alpha(t)=\alpha_0 t^a$, ${\cal D}(t)=D_0$, and $k_2(t) =0$,
where $c_1=2+N(1-q)>0$.  \label{table3}}
\begin{ruledtabular}
\begin{tabular}{cccc}
  $\alpha_0$ & $a$ & $\sigma^2 (t) $  & description \\ \hline \hline
  $(q-1)\alpha_0<0$ & $a=0$  & $\left(\frac{e^{(q-1)\alpha_0 t}-1}{(q-1)\alpha_0}\right)^{2/c_1} $
 & stationary  \\\hline
  $(q-1)\alpha_0>0$ & $a=0$  & $\left(\frac{e^{(q-1)\alpha_0 t}-1}{(q-1)\alpha_0}\right)^{2/c_1} $
 & exponentially \\ & & & diffusive \\ \hline
$\alpha_0$ & $a=-1$ & $t^{2[(q-1)\alpha_0 -1]/c_1}$ &
$c_1$-diffusive\footnotemark[1] \\ \hline
& & &less than \\
$(q-1)\alpha_0>0$ & $-1<a<0$ & $  e^{\frac{2[(q-1)\alpha_0]
t^{1+a}]}{c_1(1+a)}}$ &  exponentially \\ & & &  diffusive
\\ \hline
  $(q-1)\alpha_0<0$ & $-1<a<0$ & $t^{2|a|/c_1}$& $c_1$-diffusive\footnotemark[2] \\ \hline
 & &  & more than \\
 $(q-1)\alpha_0>0$ & $a>0$  & $  e^{\frac{2[(q-1)\alpha_0 ]t^{1+a}]}{c_1(1+a)}}$
& exponentially  \\ & & & diffusive \\ \hline
 $(q-1)\alpha_0<0$ & $a>0$ & $t^{-2a/c_1}$ & localized \\ \hline
  $\alpha_0 $ & $a<-1$ & $t^{2/c_1}$ & $c_1$-diffusive\footnotemark[3] \\
\end{tabular}
\end{ruledtabular}
\footnotetext[1]{The process is superdiffusive for
$c_1<2[(q-1)\alpha_0 -1]$, normal for $c_1=2[(q-1)\alpha_0 -1]$,
and subdiffusive for $c_1>2[(q-1)\alpha_0 -1]$.}
\footnotetext[2]{The process is superdiffusive for $c_1<2 |a|$,
normal for $c_1=2 |a|$, and subdiffusive for $c_1>2 |a|$.}
\footnotetext[3]{The process is superdiffusive for $c_1<2$, normal
for $c_1=2$, and subdiffusive for $c_1>2$.}
\end{table}



Summing up, we have investigated an $N$-dimensional nonlinear
Fokker-Planck equation by incorporating the time dependence every
coefficient, including those of the external force and the source
term. An exact solution is obtained in the case of external force
with isotropic spatial linear term and a possible anisotropic
spatial constant term. This anisotropic constant term is useful,
for instance, to analyze diffusion when the gravitational field is
relevant. We have showed that a rich class of anomalous behaviors
arises by choosing appropriate time dependence of the
coefficients. These results indicate that the anomaly in diffusive
processes may appear as a consequence of different causes. In
particular, the combination of nonlinearity and time dependence of
the coefficients can lead to the normal diffusion (time linear
growth of the mean square displacement). This fact implies that
the normal diffusion, in general, can not be associated with the
Gaussian shape of the distribution $\rho ({\bf r},t)$, {\it {i.
e.}}, the linear time increase of variance does not necessarily
mean that we are in the presence of ordinary diffusion.
 Finally, we hope that the results obtained here may be
useful to clarify a possible origin of a large class of different
anomalous diffusive processes in theoretical and experimental
contexts.

We thank CAPES, CNPq and PRONEX (Brazilian agencies)  for partial
financial support.



\begin{thebibliography}{99}
\bibitem{Goldenfeld} The connection
between renormalization group theory and Tsallis statistics,
suggested by Eq. (\ref{S2}),  is an interesting theme to be
clarified. However, this particular subject seems non-trivial and
deserves further investigations that should be addressed in
another opportunity.
\end{thebibliography}
\end{document}